# Direct writing of individual quantum dots

**Authors**
Weikun Zhu[1,2†], Natalie Ngoh[2,3†], Shelly Ben-David[2,3†], Maxwell Conte[2,4], Teddy Hsieh[2,3], Sarah O. Spector[2,3], Tara Sverko[5], Patricia Jastrzebska-Perfect[2,3], Will Jack[3], Jinwoo Sim[2,3], Peter F. Satterthwaite[2,3], Farnaz Niroui[2,3*]

**Affiliations**
[1]Department of Chemical Engineering, Massachusetts Institute of Technology, Cambridge, MA 02139, USA
[2]Research Laboratory of Electronics, Massachusetts Institute of Technology, Cambridge, MA 02139, USA
[3]Department of Electrical Engineering and Computer Science, Massachusetts Institute of Technology, Cambridge, MA 02139, USA
[4]Department of Materials Science and Engineering, Massachusetts Institute of Technology, Cambridge, MA 02139, USA
[5]Department of Chemistry, Massachusetts Institute of Technology, Cambridge, MA 02139, USA

[†]These authors contributed equally to this work.
[*]Corresponding author. Email: fniroui@mit.edu

Quantum light sources capable of generating single photons are fundamental building blocks for photonic quantum technologies[1–4]. In the ongoing search for an ideal quantum emitter, inorganic halide perovskite nanocrystals have emerged as a promising source of single photons[5–10]. Their unique optical response, with an unmatched ease of synthetic tunability[11,12], stands out amongst the competing platforms. However, their stochastic dispersion in solution challenges the deterministic and stable integration of individual emitters with photonic structures that is required for practical technologies[13]. Notably, resolution and material compatibility constraints make conventional top-down fabrication processes insufficient for such heterogeneous integration. Here, we report direct writing of perovskite quantum dots (QDs) with individual-emitter resolution. By inducing a nanoscale-confined formation volume using a thermal scanning probe method, we achieve site-selective synthesis down to a single atomic-scale QD with spectral tunability and < 25 nm spatial control. As a result, we demonstrate high-yield arrays of $CsPbI_3$ single-photon emitters with narrow linewidths and high single-photon purity up to 98% at room temperature, performance consistent with that of their state-of-the-art colloidal counterparts. Through such deterministic control, we uniquely realize the precise, on-demand coupling of these emitters to photonic cavities, as evidenced by a measured enhancement in the spontaneous emission rate. This represents a key advancement toward addressing the longstanding integration obstacles of these materials. Overall, by combining the atomic-scale tunability of chemical synthesis with the spatial control of additive manufacturing, our work opens new emitter engineering strategies to realize the untapped potential of colloidal materials for next-generation quantum technologies.

Future applications of photonic quantum technologies in computing, communication, and sensing rely on single-photon emitters[1,2]. While an ideal emitter that meets all the performance and integration needs remains elusive, diverse solid-state platforms have emerged as promising[3,4].

These include atom-like epitaxial[14] and colloidal semiconductor QDs[5,6,15], defects in diamond[16], Si[17], SiC[18], GaN[19], and carbon nanotubes[20], and trapped excitons in two-dimensional materials[21]. Amongst these, colloidal QDs, which are nanocrystals made in solution, stand out for their scalable and low-cost processing with high-precision synthetic control over size, shape, composition and surfaces, leading to a widely tunable electronic and optical response[6,12,22]. This exceptional emission tunability is accompanied by narrow spectral linewidths and high photoluminescence quantum yields[11,23]. Recent advances in QDs of halide perovskites ($CsPbX_3$, where X = Cl, Br, I) present these attributes while also uniquely realizing fast radiative lifetimes comparable to their coherence times at cryogenic temperatures[7,8,10], enabling highly pure[10,24–26] and coherent single-photon emission[7,8]. These advancements have positioned perovskite QDs as a leading colloidal source of single photons.

Their nanometer size and stochastic dispersion in solution, though, challenge control of individual QDs and integration with photonic structures, which is required for practical applications. Processing is further constrained by material instabilities that cause incompatibility with conventional top-down fabrication methods. Most state-of-the-art demonstrations, thus, involve spin-coating a diluted solution of QDs onto a surface and then selecting the desired emitter from a film of randomly distributed ones. In this approach, the lack of deterministic placement restricts functional integration, limiting applications to basic research. Several techniques have been proposed to address spatial control, including dielectrophoretic[27] or optical trapping[28], electrostatic[29] or capillary[30] assembly, DNA origami[31] and electrohydrodynamic printing[32]. However, these methods face challenges including limited resolution, low yield, incompatibility with diverse systems, and processing-induced damage to material quality and stability. As a result, achieving single QD control – essential for quantum applications – remains a significant barrier to unlocking the full potential of colloidal emitters.

Here, we introduce an alternative approach in which the QDs are directly synthesized on the substrate at the desired site with resolution down to a single emitter. Previous efforts in site-selective growth have been limited to larger perovskite crystals[33–36]. Laser-directed in-situ formation of stable QDs has also been demonstrated[37–40], however, constrained by the diffraction limit, they can only yield nanocrystal ensembles with printed features that are larger than 400 nm[39]. In our direct-write platform, we push beyond the conventional resolution limits, for the first time reaching individual quantum emitters using a confined growth volume with controlled formation induced by a thermal scanning probe in a matrix of precursor elements. As a result, we demonstrate deterministic arrays of room-temperature, high-purity $CsPbI_3$ single-photon emitters with performance competitive with their colloidal counterparts. Using their deterministic placement, we show that these emitters can be uniquely integrated on demand with photonic structures with < 25 nm accuracy, addressing their longstanding integration obstacles. Importantly, our results demonstrate that additive manufacturing with atomic-scale control of chemical synthesis is a promising solution to bridge the compatibility and resolution gaps in existing fabrication techniques to facilitate applications of emerging low-dimensional quantum materials.

**Direct writing of perovskite QDs with single-emitter resolution**

The growth medium for our site-selective synthesis platform was designed in the form of a thin polymeric matrix containing dispersed precursor elements. As a prototypical system, we used a mixture of poly(methyl methacrylate) (PMMA), CsX and $PbX_2$, where X = Cl, Br, I or their combination, dissolved in dimethylformamide and spin-coated on the substrate. Formation of the QDs in such a matrix can be induced thermally. To ensure localized growth, we used a thermal scanning probe[41,42] to apply the heat (Fig. 1a). Once in contact with the matrix, the heated probe generates a nanoscale-confined thermal volume. In the regions where the temperature approaches the polymer glass transition temperature ($T_g$), the embedded precursor ions become mobilized and facilitate nucleation and growth of the QDs. Using this approach, QDs with photoluminescence (PL) emission spanning a broad wavelength range – from 492 nm to 658 nm – were successfully synthesized by tuning the halide composition (Fig. 1b). Additionally, by spatially scanning the probe, we achieved high-resolution patterned growth in complex and arbitrary designs. An example based on $CsPbI_3$ QDs is shown in Fig. 1c.

The highly localized growth volume, combined with engineering of the precursor matrix, can allow for the site-selective growth down to an individual emitter. An array of such individual $CsPbI_3$ emitters is shown in Fig. 1d. The inset cross-sectional transmission electron microscope (TEM) image depicts a resulting QD, revealing explicit lattice fringes with a spacing of 0.31 nm, consistent with the (202) plane of $\gamma$-$CsPbI_3$. This crystalline nature of the resulting QDs is further verified by X-ray diffraction (XRD) measurements on a uniformly heated precursor matrix, where the peak at 28.9° can be assigned to the (202) plane (Extended Data Fig. 1a). With a size of ~ 4.5 nm, which is smaller than the 12 nm Bohr diameter of $CsPbI_3$,[11] these strongly quantum-confined dots are expected to display high single-photon purity[25].

The response of a representative emitter is shown in Fig. 1f – h. This emitter exhibited a room-temperature PL emission peak centered at 665 nm with a narrow linewidth (full-width at half-maximum) of 67 meV (Fig. 1f), and a PL lifetime of 13 ns (Extended Data Fig. 1b). The intensity time trace revealed characteristic blinking behavior between bright and dark states, indicative of single QD emission (Fig. 1g). Notably, the emitter remained predominantly in its bright emissive state, with an on-time fraction of 94%. Single photon emission was verified using a Hanbury Brown and Twiss (HBT) photon correlation measurement, which demonstrated a pronounced antibunching characterized by a second-order intensity correlation value at zero time delay, $g^{(2)}(0)$, of 0.04, well below the 0.5 threshold for single-photon emission (Fig. 1h). This confirms the presence of a single emitter at the targeted site, exhibiting a high single-photon purity at room temperature. Compared to the best-reported colloidal $CsPbI_3$ QDs[24–26,43–45], our on-site synthesized QDs deliver competitive performance while uniquely enabling deterministic, site-selective engineering and integration.

**Working principle towards high-purity single-photon emission**

The high resolution of our approach stems in part from the nanoscale localization offered by the thermal probe. Fig. 2a shows finite element method simulation of the localized heating for a silicon probe with a 5 nm tip radius, heated to 850 °C and indented 15 nm into a 25 nm-thick precursor film. As shown, the set temperature drops notably before reaching the sample due to thermal dissipation and imperfect heat transfer at the probe-sample interface[42]. The resulting thermal profile in the precursor film is additionally influenced by the thermal conductivities of both the film and the substrate, as well as the film's thickness. Within this highly localized thermal volume, QD formation is confined to regions where the temperature is sufficient to facilitate growth while avoiding any degradation. In the case of the 25 nm-thick PMMA matrix, this optimal temperature window was characterized to be ~ 110 °C – 140 °C (Supplementary Note 1), representing a small fraction of the overall hotspot, thereby further focusing the growth volume and enhancing the resolution.

The onset temperature of growth depends on the matrix material. As the matrix undergoes the glass-transition temperature, it transforms from a glassy to a rubbery state where polymer chains gain segmental mobility[46]. This generates a volume in which the embedded precursor ions can freely diffuse, promoting a localized nucleation and growth of the QDs. To validate this concept, we tested the QD formation in a series of precursor films made with different $T_g$ values by blending PMMA with a low-$T_g$ polymer, polyvinylidene fluoride (PVDF). As the PVDF weight ratio was increased from 0% to 50%, the $T_g$ of the blend dropped from 124 °C to 61 °C (Extended Data Fig. 2). This change in $T_g$ was directly correlated with a shift in the measured onset temperature of growth as shown in Fig. 2b-c, confirming the critical role that the matrix plays in defining the QD growth process and its window. It should be noted that the matrix additionally serves as an encapsulant to help protect the emitter from ambient exposure and minimize potential degradation[47].

To optimize the writing process for individual-emitter resolution, the amount of precursor ions available for growth must be tightly constrained. We achieved this through simultaneously reducing the precursor concentration and the growth volume. The latter can be modified using the precursor film thickness and the applied temperature. Given the suppressed biexciton quantum yield expected in these highly confined QDs[25,48], we used $g^{(2)}(0) < 0.5$ as an indicator of reaching individual emitters. As shown in Fig. 2d, $g^{(2)}(0)$ increased for higher precursor concentrations, accompanied by the emergence of broader and multi-peak emission in the PL response, reflected in the increasing effective linewidth (Supplementary Note 2). Together, these suggest the formation of multiple emitters. Reducing the precursor film thickness further assists formation of individual QDs (Fig. 2e) by minimizing the growth volume (Supplementary Note 3 and Supplementary Fig. 3a-b). The applied temperature also plays a role, resulting in the trend observed in Fig. 2f, which is influenced by multiple competing effects. Specifically, we expect a combination of a temperature-dependent increase in thermal volume and an asymmetry in the growth profile influenced by the substrate[49,50] (Supplementary Fig. 3c). Through these experiments, we observed optimal growth of high-purity single-photon emitters in 25 nm-thick PMMA film with 10.3 wt% precursors and patterned at an applied probe temperature of 850 °C.

Patterning was completed with > 85% yield of optically active sites, with higher yields feasible through tuning the writing conditions (Extended Data Fig. 3). The PL response of 65 such $CsPbI_3$ single-emitter sites is presented in Fig. 3a. These emitters exhibited a consistent response, with a peak emission centered at 645.3 ± 13.9 nm (Fig. 3b), a linewidth of 92.3 ± 21.7 meV (Fig. 3c), and a lifetime of 11.0 ± 2.7 ns (Fig. 3d), aligning well with characteristics of their colloidal counterparts[24–26,43–45]. Notably, 98% of these sites demonstrated single-photon emission, confirmed by a measured $g^{(2)}(0) < 0.5$. The $g^{(2)}(0)$ statistical distribution is shown in Fig. 3e with an average value of 0.12 ± 0.11, and a measured high single-photon purity reaching up to 98% at room temperature. Our direct-write emitters displayed stable emission (Extended Data Fig. 4) with a representative source-level brightness of ~ 11% (Supplementary Note 4). Supplementary Table 1 benchmarks the QDs in this work against their colloidal counterparts.

## Deterministic integration with photonic structures

The spatial control of our direct-write approach enables the emitters to be deterministically integrated with photonic structures – a feature required for controlling and enhancing their emission characteristics towards functional applications. We demonstrated this capability using a $TiO_2$ dielectric circular Bragg grating (CBG) microcavity, consisting of a central disk surrounded by concentric Bragg rings (Fig. 4a). As shown in Supplementary Fig. 5a, the electric field is tightly confined at the cavity's center, where the increased local density of optical states leads to enhancement of the radiative rate for an emitter placed at this location, provided its emission overlaps spectrally with the optical resonance mode. We numerically optimized the cavity design for our $CsPbI_3$ QDs using finite-difference time-domain (FDTD) simulations, followed by experimental fine-tuning as detailed in Supplementary Note 5. The finalized design, exhibiting a resonance at 640 nm, was fabricated with the structure shown in the scanning electron microscope (SEM) image of Fig. 4b.

For optimal performance and maximal coupling effect, the emitter needs to be positioned within ~ 50 nm of the cavity center (Extended Data Fig. 5a-b). Indeed, using our approach, we achieved direct growth of the emitter with < 25 nm placement accuracy as characterized in Extended Data Fig. 5c-d for over 60 samples. An example of a $CsPbI_3$ QD precisely coupled to the center of a CBG cavity is presented in Fig. 4c. The PL spectrum of this emitter displays the expected cavity-coupled peak at 640 nm, with a quality factor (Q) of ~ 200 (Fig. 4d). The coupling was further confirmed by a measured reduction in the QD emission decay time to 3.7 ns (Fig 4e) compared to the decay on a planar $SiO_2$ substrate (11.0 ± 2.7 ns) – corresponding to a ~ 3-fold enhancement in the spontaneous emission rate. Notably, the emitter maintained its high-purity single-photon emission through the coupling process, as evidenced by its low $g^{(2)}(0) = 0.04$ (Fig. 4f). Such deterministic control helps address the persistent integration challenge faced by stochastic colloidal systems, thereby accelerating research and enabling new opportunities to realize applications of perovskite emitters.

## Conclusion

In summary, we present a platform for direct-writing of perovskite QDs, enabled by a thermally-induced, site-selective synthesis, achieving spatial patterning of complex designs with tunable optical response and resolution down to the single-emitter level. Leveraging this capability, we demonstrated deterministic arrays of high-purity $CsPbI_3$ single-photon emitters with precise integration into photonic cavities, realizing < 25 nm spatial control. This unique ability to generate and position perovskite emitters on demand and with spatial precision addresses the longstanding integration challenges of these materials, promoting their growing potential in quantum technologies. Notably, the implications extend beyond perovskites. The expanding class of low-dimensional quantum materials broadly faces compatibility and resolution constraints inherent to top-down fabrication techniques. Our work showcases additive manufacturing, where atomic-scale precision of bottom-up chemical synthesis is combined with spatial control, as a promising path to bridge this technology gap. Thus, enabling these materials to advance into next-generation devices and systems with cross-disciplinary impact on emerging areas of computing, communication, sensing and information processing.

## Data availability

The data that support the findings of this study are available within the paper and its Supplementary Information. All other relevant data are available from the corresponding author upon reasonable request.

## Acknowledgements

This work was supported by the National Science Foundation (NSF) Award DMR-2144136 and the MIT Center for Quantum Engineering (CQE). N.N. acknowledges support by the Singapore National Science Scholarship from the Agency for Science, Technology and Research (A*STAR). S.B.D. acknowledges support from the MIT George and Marie Vergottis Fellowship. P.J.P., S.O.S., P.F.S. and T.H. acknowledge support from the NSF Graduate Research Fellowship Program under Grant No. 1745302. The fabrication and characterization procedures in this work have been in part carried out using the MIT.nano shared facilities. The TEM sample preparation and imaging were performed at the Harvard University Center for Nanoscale Systems (CNS), a member of the National Nanotechnology Coordinated Infrastructure Network (NNCI), which is supported by the National Science Foundation under NSF ECCS Award No. 1541959. The authors thank Heidelberg Instruments for technical support.

## Author contributions

F.N. conceived of the project and supervised its implementation. W.Z., N.N., S.B.D. and F.N. designed and performed the experiments, and wrote the manuscript. W.Z. analyzed and plotted the results. M.C. performed the XRD, XPS, and glass transition temperature measurements and analysis. T.S. and P.J.P. assisted with the development of the optical characterization setup. W.J. helped with the initial process development of the direct-write platform. T.H., S.O.S., J.S. and P.F.S. assisted with developing the fabrication process for the optical cavities. All authors contributed to finalizing the manuscript.

## Competing interests

A patent application has been filed with the United States Patent and Trademark Office, covering the work presented in this paper.

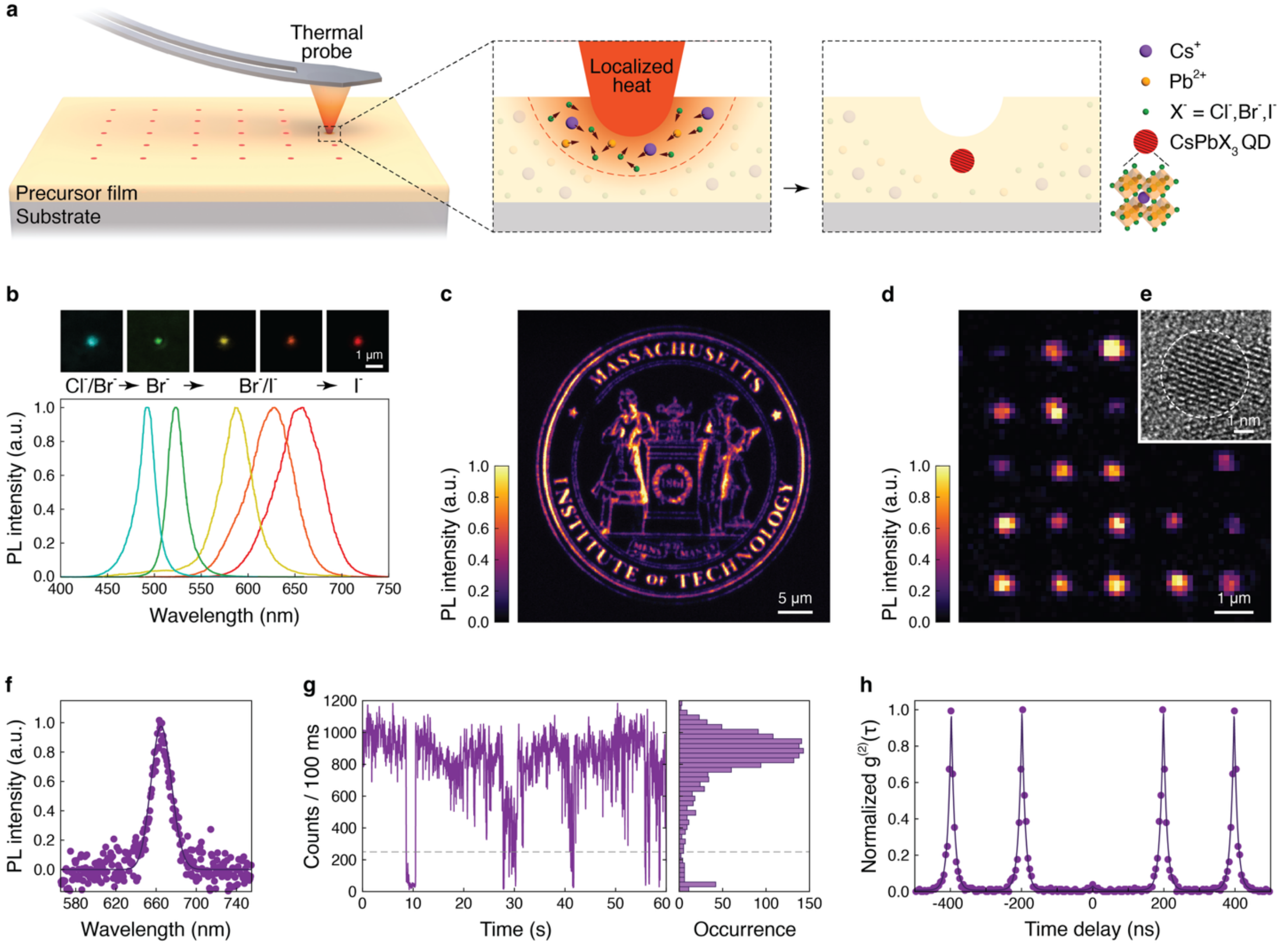

**Fig. 1 | Direct writing of perovskite QDs with single-emitter resolution. a,** Schematic illustration of the direct-write approach where a thermal probe generates a nanoscale localized growth volume transforming precursors embedded in a matrix into perovskite QDs. **b,** PL map and spectra of patterned QDs with PL emission tuned across a broad range by using different halide compositions. Scale bar, 1 μm. **c,** PL map of a high-resolution pattern produced with $CsPbI_3$ QDs. Scale bar, 5 μm. **d,** PL map of an array of $CsPbI_3$ single emitters. Scale bar, 1 μm. **e,** TEM image of a direct-write $CsPbI_3$ QD. Scale bar, 1 nm. **f,** PL spectrum of a representative single-emitter site with emission peak at 665 nm and 67 meV linewidth. **g,** Intensity time trace of this emitter with on-state fraction of 94%. The dashed line represents the on-state threshold. **h,** Second-order photon correlation function, $g^{(2)}(\tau)$, of the emitter showing $g^{(2)}(0) = 0.04$, confirming high-purity single-photon emission at room temperature. All the results correspond to samples fabricated on $SiO_2$-on-Si substrates.

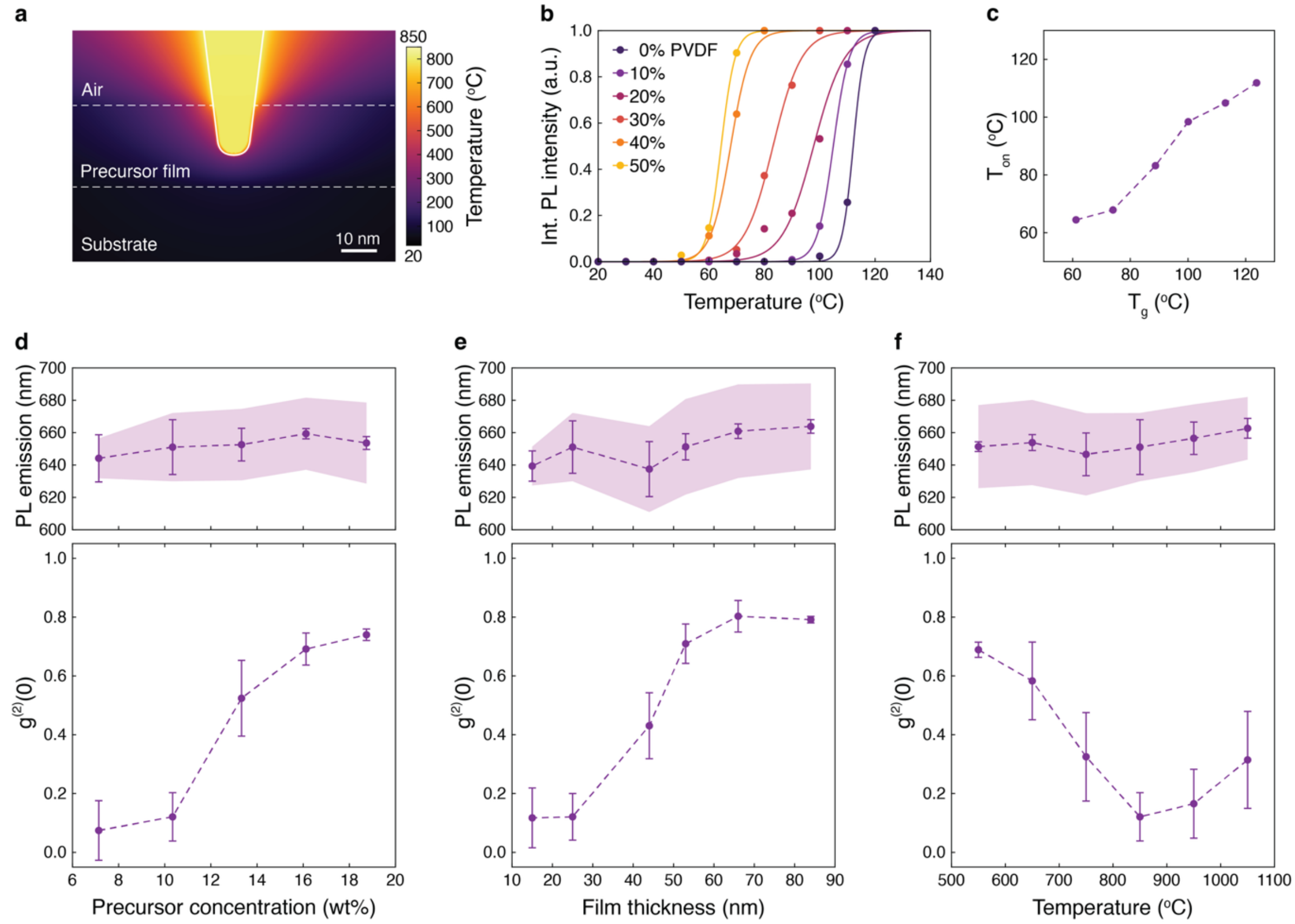

**Fig. 2 | Direct writing optimization for single-photon emitters. a,** Simulated thermal profile of a Si probe heated to 850 °C indented 15 nm into a 25 nm-thick PMMA matrix on a $SiO_2$-on-Si substrate. Scale bar, 10 nm. **b,** $CsPbI_3$ QD growth evolution in a PMMA matrix with increasing weight-ratio of PVDF from 0% to 50% under uniform heating. The plot shows the integrated PL intensity collected for samples formed at increasing temperatures, following the discussion in Supplementary Note 1. Increase in PVDF lowers $T_g$ of the blend (Extended Data Fig. 2), which corresponds to a decrease in the onset temperature of growth ($T_{on}$) defined as the inflection point of the fitted sigmoid. **c,** $T_{on}$-dependence on $T_g$ of the precursor matrix, extracted from Fig. 2b and Extended Data Fig. 2. **d-f,** The process can be optimized for writing individual emitters by tuning the precursor concentration **(d)**, film thickness **(e)**, and applied probe temperature **(f)**, characterized by minimizing $g^{(2)}(0)$ to realize high-purity single-photon emission. Top panels show the PL peak emission and the effective linewidth (shaded region) as a function of the change in the growth conditions. Effective linewidth is described in Supplementary Note 2. Data are presented as mean ± standard deviation for over 20 sites per condition. These samples are prepared in a PMMA matrix.

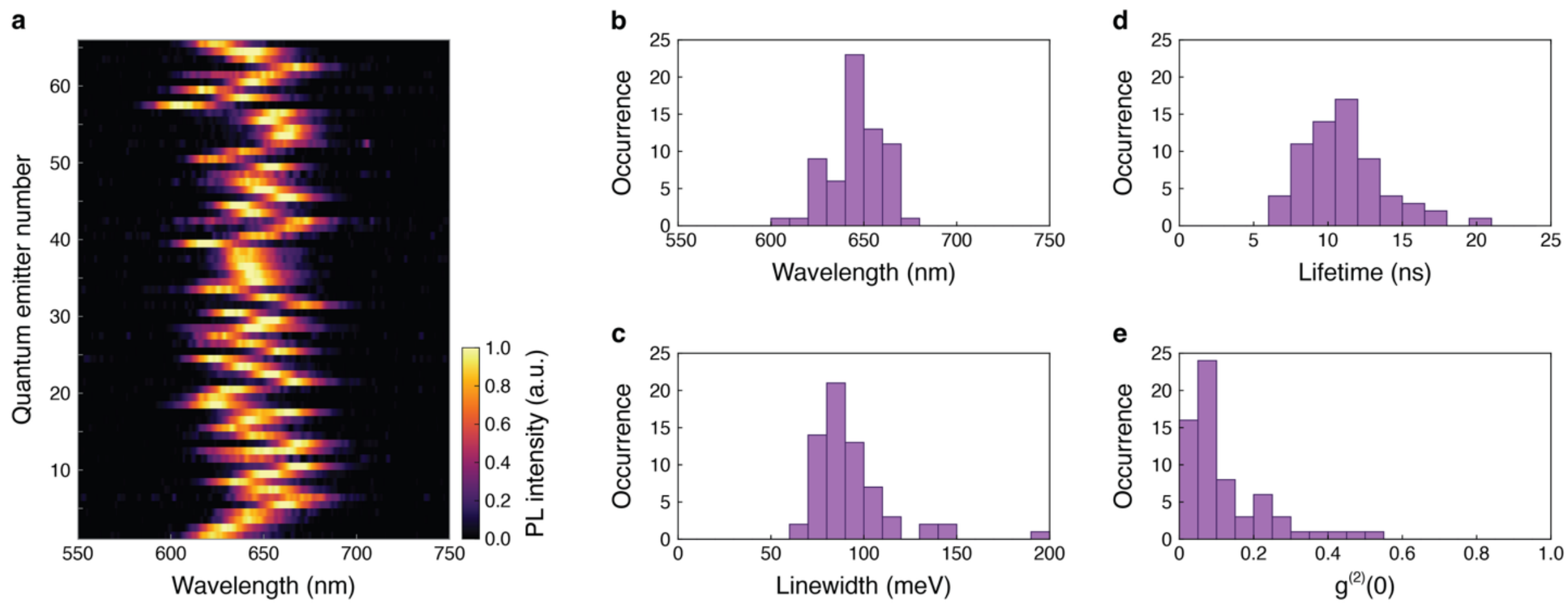

**Fig. 3 | Optical characteristics of direct-write $CsPbI_3$ single-photon emitters. a,** PL spectra of 65 single-emitter sites patterned in a 25 nm-thick PMMA film with 10.3 wt% precursors at an applied probe temperature of 850 °C. **b-e,** Distribution of the peak PL emission **(b)**, linewidth **(c)**, lifetime **(d)**, and $g^{(2)}(0)$ **(e)** of the emitters in **(a)**.

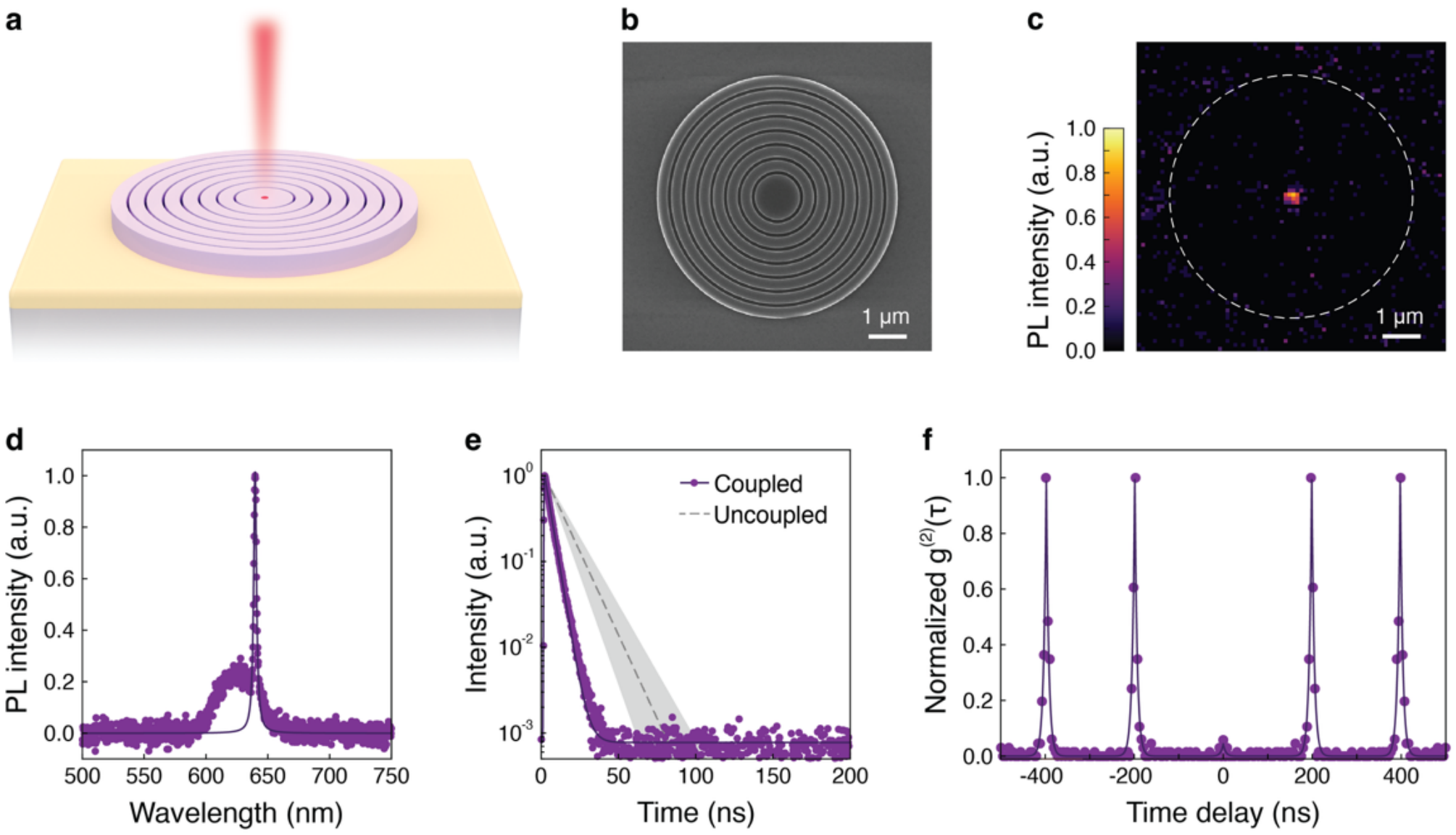

**Fig. 4 | Deterministic coupling of the direct-write QDs to photonic cavities. a,** Schematic of a QD integrated with a circular Bragg grating (CBG) cavity. **b,** SEM image of a fabricated $TiO_2$ CBG cavity with a resonance at 640 nm. **c,** PL map of a $CsPbI_3$ QD directly grown at the center of the cavity. **d,** PL spectrum of this QD showing enhanced emission at the cavity resonance. **e,** Time-resolved PL decay of this emitter showing ~ 3-fold enhancement in the spontaneous emission rate compared to the emitters on planar $SiO_2$ plotted as mean (dotted line) ± standard deviation (shaded region) based on the results in Fig. 3d. **f,** Second-order photon correlation function, $g^{(2)}(\tau)$, of this emitter showing $g^{(2)}(0) = 0.04$, confirming high-purity single-photon emission when coupled to the cavity.

## References

1. O'Brien, J. L., Furusawa, A. & Vučković, J. Photonic quantum technologies. *Nat. Photonics* **3**, 687–695 (2009).
2. Couteau, C. *et al.* Applications of single photons to quantum communication and computing. *Nat. Rev. Phys.* **5**, 326–338 (2023).
3. Aharonovich, I., Englund, D. & Toth, M. Solid-state single-photon emitters. *Nat. Photonics* **10**, 631–641 (2016).
4. Esmann, M., Wein, S. C. & Antón-Solanas, C. Solid-state single-photon sources: Recent advances for novel quantum materials. *Adv. Funct. Mater.* **34**, 2315936 (2024).
5. Zhu, J., Li, Y., Lin, X., Han, Y. & Wu, K. Coherent phenomena and dynamics of lead halide perovskite nanocrystals for quantum information technologies. *Nat. Mater.* **23**, 1027–1040 (2024).
6. Almutlaq, J. *et al.* Engineering colloidal semiconductor nanocrystals for quantum information processing. *Nat. Nanotechnol.* **19**, 1091–1100 (2024).
7. Utzat, H. *et al.* Coherent single-photon emission from colloidal lead halide perovskite quantum dots. *Science* **363**, 1068–1072 (2019).
8. Kaplan, A. E. K. *et al.* Hong–Ou–Mandel interference in colloidal $CsPbBr_3$ perovskite nanocrystals. *Nat. Photonics* **17**, 775–780 (2023).
9. Zhu, C. *et al.* Single-photon superradiance in individual caesium lead halide quantum dots. *Nature* **626**, 535–541 (2024).
10. Jun, S. *et al.* Ultrafast and bright quantum emitters from the cavity-coupled single perovskite nanocrystals. *ACS Nano* **18**, 1396–1403 (2024).
11. Protesescu, L. *et al.* Nanocrystals of cesium lead halide perovskites ($CsPbX_3$, X = Cl, Br, and I): Novel optoelectronic materials showing bright emission with wide color gamut. *Nano Lett.* **15**, 3692–3696 (2015).
12. Shamsi, J., Urban, A. S., Imran, M., De Trizio, L. & Manna, L. Metal halide perovskite nanocrystals: Synthesis, post-synthesis modifications, and their optical properties. *Chem. Rev.* **119**, 3296–3348 (2019).
13. Lee, J. *et al.* Integrated single photon emitters. *AVS Quantum Sci.* **2**, 031701 (2020).
14. Senellart, P., Solomon, G. & White, A. High-performance semiconductor quantum-dot single-photon sources. *Nat. Nanotechnol.* **12**, 1026–1039 (2017).
15. Nelson, D., Byun, S., Bullock, J., Crozier, K. B. & Kim, S. Colloidal quantum dots as single photon sources. *J. Mater. Chem. C* **12**, 5684–5695 (2024).
16. Aharonovich, I. *et al.* Diamond-based single-photon emitters. *Rep. Prog. Phys.* **74**, 076501 (2011).
17. Redjem, W. *et al.* All-silicon quantum light source by embedding an atomic emissive center in a nanophotonic cavity. *Nat. Commun.* **14**, 3321 (2023).
18. Castelletto, S. Silicon carbide single-photon sources: challenges and prospects. *Mater. Quantum Technol.* **1**, 023001 (2021).
19. Berhane, A. M. *et al.* Bright room-temperature single-photon emission from defects in gallium nitride. *Adv. Mater.* **29**, 1605092 (2017).
20. Ma, X., Hartmann, N. F., Baldwin, J. K. S., Doorn, S. K. & Htoon, H. Room-temperature single-photon generation from solitary dopants of carbon nanotubes. *Nat. Nanotechnol.* **10**, 671–675 (2015).

21. Turunen, M. *et al.* Quantum photonics with layered 2D materials. *Nat. Rev. Phys.* **4**, 219–236 (2022).
22. García De Arquer, F. P. *et al.* Semiconductor quantum dots: Technological progress and future challenges. *Science* **373**, eaaz8541 (2021).
23. Dey, A. *et al.* State of the art and prospects for halide perovskite nanocrystals. *ACS Nano* **15**, 10775–10981 (2021).
24. Park, Y.-S., Guo, S., Makarov, N. S. & Klimov, V. I. Room temperature single-photon emission from individual perovskite quantum dots. *ACS Nano* **9**, 10386–10393 (2015).
25. Zhu, C. *et al.* Room-temperature, highly pure single-photon sources from all-inorganic lead halide perovskite quantum dots. *Nano Lett.* **22**, 3751–3760 (2022).
26. Mi, C. *et al.* Towards non-blinking and photostable perovskite quantum dots. *Nat. Commun.* **16**, 204 (2025).
27. Barik, A., Chen, X. & Oh, S.-H. Ultralow-power electronic trapping of nanoparticles with sub-10 nm gold nanogap electrodes. *Nano Lett.* **16**, 6317–6324 (2016).
28. Jensen, R. A. *et al.* Optical trapping and two-photon excitation of colloidal quantum dots using bowtie apertures. *ACS Photonics* **3**, 423–427 (2016).
29. Xing, X. *et al.* High-resolution combinatorial patterning of functional nanoparticles. *Nat. Commun.* **11**, 6002 (2020).
30. Barelli, M. *et al.* Single-photon emitting arrays by capillary assembly of colloidal semiconductor CdSe/CdS/$SiO_2$ nanocrystals. *ACS Photonics* **10**, 1662–1670 (2023).
31. Chen, C. *et al.* Ultrafast dense DNA functionalization of quantum dots and rods for scalable 2D array fabrication with nanoscale precision. *Sci. Adv.* **9**, eadh8508 (2023).
32. Guymon, G. G. *et al.* Deterministic printing and heterointegration of single quantum dots. arXiv:2501.04177.
33. Jastrzebska-Perfect, P. *et al.* On-site growth of perovskite nanocrystal arrays for integrated nanodevices. *Nat. Commun.* **14**, 3883 (2023).
34. Du, J. S. *et al.* Halide perovskite nanocrystal arrays: Multiplexed synthesis and size-dependent emission. *Sci. Adv.* **6**, eabc4959 (2020).
35. Lin, C.-K. *et al.* Two-step patterning of scalable all-inorganic halide perovskite arrays. *ACS Nano* **14**, 3500–3508 (2020).
36. Chen, X.-G. *et al.* Optofluidic crystallithography for directed growth of single-crystalline halide perovskites. *Nat. Commun.* **15**, 3677 (2024).
37. Sun, K. *et al.* Three-dimensional direct lithography of stable perovskite nanocrystals in glass. *Science* **375**, 307–310 (2022).
38. Huang, X. *et al.* Reversible 3D laser printing of perovskite quantum dots inside a transparent medium. *Nat. Photonics* **14**, 82–88 (2020).
39. Hu, J. *et al.* Full-color and high-resolution femtosecond laser patterning of perovskite quantum dots in polyacrylonitrile matrix. *Adv. Funct. Mater.* **34**, 2407116 (2024).
40. Zhan, W. *et al.* In situ patterning perovskite quantum dots by direct laser writing fabrication. *ACS Photonics* **8**, 765–770 (2021).
41. Albisetti, E. *et al.* Thermal scanning probe lithography. *Nat. Rev. Methods Primer* **2**, 32 (2022).
42. Howell, S. T., Grushina, A., Holzner, F. & Brugger, J. Thermal scanning probe lithography—a review. *Microsyst. Nanoeng.* **6**, 21 (2020).
43. Farrow, T. *et al.* Ultranarrow line width room-temperature single-photon source from perovskite quantum dot embedded in optical microcavity. *Nano Lett.* **23**, 10667–10673 (2023).

44. Hsu, B.-W. *et al.* Very robust spray-synthesized $CsPbI_3$ quantum emitters with ultrahigh room-temperature cavity-free brightness and self-healing ability. *ACS Nano* **15**, 11358–11368 (2021).
45. Hu, F. *et al.* Slow Auger recombination of charged excitons in nonblinking perovskite nanocrystals without spectral diffusion. *Nano Lett.* **16**, 6425–6430 (2016).
46. Hall, D. B. & Torkelson, J. M. Small molecule probe diffusion in thin and ultrathin supported polymer films. *Macromolecules* **31**, 8817–8825 (1998).
47. Raja, S. N. *et al.* Encapsulation of perovskite nanocrystals into macroscale polymer matrices: Enhanced stability and polarization. *ACS Appl. Mater. Interfaces* **8**, 35523–35533 (2016).
48. Li, Y., Luo, X., Ding, T., Lu, X. & Wu, K. Size- and halide-dependent Auger recombination in lead halide perovskite nanocrystals. *Angew. Chem.* **132**, 14398–14401 (2020).
49. Xia, W., Mishra, S. & Keten, S. Substrate vs. free surface: Competing effects on the glass transition of polymer thin films. *Polymer* **54**, 5942–5951 (2013).
50. Ma, Z., Nie, H. & Tsui, O. K. C. Substrate influence on the surface glass transition temperature of polymers. *Polymer* **312**, 127594 (2024).
51. Yamasue, E., Susa, M., Fukuyama, H. & Nagata, K. Thermal conductivities of silicon and germanium in solid and liquid states measured by non-stationary hot wire method with silica coated probe. *J. Cryst. Growth* **234**, 121–131 (2002).
52. Grove, A. S. *Physics and Technology of Semiconductor Devices*. (Wiley, 1967).
53. Assael, M. J., Botsios, S., Gialou, K. & Metaxa, I. N. Thermal conductivity of polymethyl methacrylate (PMMA) and borosilicate crown glass BK7. *Int. J. Thermophys.* **26**, 1595–1605 (2005).
54. Losego, M. D., Moh, L., Arpin, K. A., Cahill, D. G. & Braun, P. V. Interfacial thermal conductance in spun-cast polymer films and polymer brushes. *Appl. Phys. Lett.* **97**, 011908 (2010).